%% file: ms.tex
\documentclass[manuscript]{aastex}

\shorttitle{The Bar Pattern Speed of NGC 1433}
\shortauthors{Treuthardt, Salo, Rautiainen, \& Buta}

\begin{document}


\title{The Bar Pattern Speed of NGC 1433 Estimated Via Sticky-Particle Simulations}


\author{P. Treuthardt\altaffilmark{1},  H. Salo\altaffilmark{2}, P.
Rautiainen\altaffilmark{2}, and  R. Buta\altaffilmark{1}}
\altaffiltext{1}{Department of Physics and Astronomy, University of Alabama,
Box 870324, Tuscaloosa, AL 35487, USA}\altaffiltext{2}{Division of Astronomy,
Department of Physical Sciences, University of Oulu, Oulu, FIN-90014, Finland}



\begin{abstract}
We present detailed numerical simulations of NGC 1433, an intermediate-type
barred spiral showing strong morphological features including a secondary bar,
nuclear ring, inner ring, outer pseudoring, and two striking, detached
spiral arcs known as ``plumes.''
This galaxy is an ideal candidate for recreating the
observed morphology through dynamical models and determining the pattern
speed. We derived a gravitational potential from an $H$-band image of the
galaxy and simulated the behavior of a two-dimensional disk of 100,000
inelastically colliding gas particles. We find that the closest matching
morphology between a $B$-band image and a simulation occurs with a
pattern speed of 0.89 km s$^{-1}$ arcsec$^{-1}$ $\pm$ 5-10\%. We also
determine that the ratio of corotation radius to the average published bar
radius is 1.7 $\pm$ 0.3, with the ambiguity in the bar radius
being the largest contributor to the error.

\end{abstract}




\section{Introduction}

NGC 1433 is a well studied disk galaxy that hosts many interesting
morphological features. Detailed studies of the galaxy have
been performed by Buta (1986) and, most recently, by Buta et al. (2001).
The galaxy is a ringed barred spiral of revised de Vaucouleurs type
(R$_1^\prime$)SB(r)ab (Buta et al. 2007; see Figure 1)
that displays a nuclear ring or lens as well as
inner and outer rings.
The nuclear ring is a small, blue feature approximately 19$\arcsec$
in angular diameter that has a small secondary bar crossing it
when viewed in the
near-infrared (Buta 1986). The strong primary bar is encircled by a
large, knotty, and intrinsically oval
inner ring which appears to be formed by
four tightly wrapped spiral arm sections that have come together in
the bar major and minor axis regions. In the context of passive,
non-self-gravitating sticky particle simulations (Schwarz 1981; 1984),
this morphology implies the
inner ring is a 4:1 resonance feature. The faint
outer pseudoring is slightly asymmetric and appears relatively
smooth compared to the inner ring.

NGC 1433 also harbors two symmetric detached arm segments, or
"plumes" (Buta 1984), just outside the leading sections of the inner
ring. Like the inner ring, the plumes lie along an ellipse oriented
parallel to the bar. It is possible that these features are residual
traces of a characteristic $m=4$ spiral pattern that may form in the
presence of bars between the inner and outer 4:1 resonances
(Rautiainen, Salo, \& Buta 2004). Since strong plumes like those seen
in NGC 1433 are rarely observed
(Buta 1984), it is possible that they are transient features although
the simulations of Rautiainen, Salo, \& Buta (2004) produced a
long-lived four-armed pattern. The plumes, as well as the nucleus and
inner ring of the galaxy, are well covered with H$\alpha$ emission (Figure 1).
The outer pseudoring is much fainter and very little H$\alpha$
emission is seen in the bar region. H I counterparts of the plumes, nuclear
ring, inner ring, and outer pseudoring have also been detected
(Ryder et al. 1996).

The pattern speed, $\Omega_p$, of a galaxy is an important parameter
that governs the overall morphology (Kalnajs 1991). In galaxies with
a single perturbation mode, such as a bar or a grand design spiral,
$\Omega_p$ determines the locations of important resonances. Since
NGC 1433 has multiple features that are thought to be associated
with specific resonances, it is an ideal candidate for recreating the
observed morphology through dynamical models (e.g., Salo et al.
1999; Rautiainen et al. 2005) and determining
$\Omega_p$. This modeling involves determining the gravitational potential
from observations and studying the "gas" response on a rigidly
rotating nonaxisymmetric component. The gas is modeled as inelastically
colliding, or "sticky," particles. Different dynamical
parameters are varied until the morphology and kinematics of the numerically
simulated galaxy matches, as closely as possible, to the morphology
and kinematics of the observed galaxy.

\section{Analysis}

\subsection{Determining the Potential}

In order to simulate the dynamics of the galaxy, a gravitational
potential must be derived. One way to do this is to assume the
near-infrared light distribution effectively traces the mass
distribution. The potential may then be derived from the convolution
of the density with the function 1/$R$ (Quillen, Frogel, \& Gonzalez
1994), combined with an assumption for the vertical density
distribution. For NGC 1433, the potential was derived from a 1.65 $\mu$m
Johnson $H$-band image obtained in February 1996 with the CTIO 1.5m
telescope (Buta et al. 2001). $B$- and $I$-band images were also taken and
the details of these observations are found in Buta et al. (2001). The
$H$-band light distribution is more reliable than $B$ or $I$ for this
purpose due to the reduced extinction effects from dust and the
weakened influence of young Population I complexes. However,
the polar FFT method we use for the evaluation of gravitational forces
(Salo et al. 1999, Laurikainen \& Salo 2002) effectively filters out
the high frequency components of the image, and is thus not sensitive
to individual SF regions (see Salo et al. 2004). Before the
gravitational potential was derived, the $H$-band image was
converted into a surface mass density by assuming a mass-to-light
($M/L$) ratio accounting for the average disk $B-H$ color gradient using
formulae from Bell \& de Jong (2001). The seven different $M/L$
models discussed in the Bell \& de Jong paper were applied to the
$H$-band image and each corrected image was used to determine a
gravitational potential (Figure 2). The $M/L$ models in Bell \& de Jong
were derived by
following the evolution of simulated exponential gaseous disks in
order to describe the trends in spiral galaxy colors.
Each $M/L$ model corresponds to the various allowed parameters in the
simulations.
The Cole et al. (2000) model discussed by Bell \& de Jong serves as a
completely independent assessment of effects of galaxy evolution on
these color trends.

The gravitational potentials were calculated for each of the seven
$M/L$-corrected images by first using an iterative two-dimensional
bulge-disk-bar decomposition described by
Laurikainen et al. (2004). Upon application of the bulge-disk-bar
decomposition it was found that the images that were $M/L$-corrected
using the outflow and Cole et al. (2000) models did not converge
through the iterative process and, therefore, were not used.
The bulge component was then removed from the $H$-band light
distribution and the disk was deprojected to a face-on orientation using
a position angle of 21$^{\circ}$ and an inclination of 33$^{\circ}$ (Buta et
al. 2001).
The disk light distribution was approximated by an azimuthal Fourier
decomposition at each radius and the disk gravity was calculated using the
even components from $m=0$ to $m=8$, as was done by Rautiainen et al.
(2005).

Other factors affect the morphology of a simulated galaxy besides
the pattern speed. These include the impact frequency from
the assumed cross section of the particles, initial particle
distribution, and time when the features are examined. The amplitude
of the bar is also an important factor affecting the modeled morphology.
Separate exponential vertical
distributions were, therefore, assumed for the
disk and the bar. The bar component was isolated from the disk
component upon determining the approximate length and, thereby, the
orientation and ellipticity of the bar in the sky plane. The
deprojected radius of the bar was already determined to be a maximum of
approximately 115$\arcsec$ by Block et al. (2004; see their Figure
1) through their Fourier bar-spiral separation technique on a deep
K$_s$-band image. The orientation and ellipticity values of the bar
at the given length in the sky plane were determined using ELLIPSE
in IRAF\footnote{IRAF is distributed by the National Optical Astronomy
Observatories,
which are operated by AURA, Inc., under cooperative agreement with the
National
Science Foundation.}.
A typical scale height of 1/6 the radial scale length (de
Grijs 1998) was assumed for the disk component of the galaxy.
Bureau et al. (2006) has advocated that the bar region of galaxies
is subject to vertical thickening. Since the thickening
factor is very uncertain, in Figure 3 we compare a range of thicknesses to
$Q_g$,
the maximum relative tangential perturbation $F_{T}/F_{R}$.
This figure shows that a fairly large range in bar scale heights
results in a rather small range of $Q_g$.
We choose the scale height of the bar to be three times that of the disk, where
$Q_g$ is approximately 0.31. This corresponds to reducing the bar amplitude
by a factor of 0.66, if the scale heights of the bar and disk are
assumed to be equal (see Figures 9 and 10 for a display of the
effects of bar amplitude variations).
In comparison with the results of Buta et al. (2005; see their Figure 8),
this value of $Q_g$ means that approximately 70$\%$ of normal bright galaxies
have
bars weaker than that seen in NGC 1433. It has been argued that boxy and
peanut shaped bulges are associated with relatively strong bars (Bureau et al.
2006) and that the ratio of bar length to thickness is 14 $\pm$ 4 for galaxies
with these bulges (L\"{u}tticke, Dettmar, \& Pohlen 2000). Therefore, we
conclude that our bar scale height of 23$\arcsec$ (corresponding to a bar
length to thickness ratio of about 10) is not unreasonable for the assumed
bar radius.

The gravitational potential of
the bulge was also added to the disk potential under the assumption
that the bulge mass is spherically distributed. Additionally, a dark
halo component, based on the universal rotation curve of Persic et
al. (1996) and scaled by approximately 1/3, was added to the galaxy
potential so that the rotation amplitude at large
radii (r $>$ 100$\arcsec$) matched the composite rotation curve of Buta et al.
(2001) as closely as possible (Figure 4). The effects of assuming a
significantly more prominent halo can be estimated from the bar amplitude
variation plots (see Figures 9 and 10). In Figure 4 there is a clear
difference between the observed rotation curve, adopted by Buta et al. (2001),
and that estimated from the
azimuthally averaged mass model from approximately 30 to 100 arcseconds.
The observed rotation curve was derived from H I, H$\alpha$, and Mg I b
data, where the H I rotation curve points were determined in Ryder et al.
(1996) using the
ROTCUR algorithm (based on the method discussed by Begeman 1989) within the
Astronomical Image Processing System (AIPS).
The H$\alpha$ rotation points were derived from Fabry-Perot velocity field
observations in Buta et al. (2001) using the Warner et al. (1973) method.
The Mg I b major axis stellar velocity points were determined using the
Fourier quotient (STSDAS routine FQUOT in IRAF) and cross-correlation methods
(STSDAS routine XCOR; see Buta et al. 2001) and then applying the asymmetric
drift correction to make them comparable with gas velocity data.
It should be noted that the observed inner rotation curve is
uncertain due to non-circular motions and the uncertainties in the
axisymmetric drift correction.

\subsection{Simulations}

The simulation code used to model NGC 1433 was written by H. Salo,
the details of which can be found in Salo et al.
(1999; see also Salo 1991 for the description of the treatment of inelastic
collisions). The behavior of a two-dimensional disk of 100,000
inelastically colliding gas particles was simulated in the
determined potential. The nonaxisymmetric part of the potential (i.e. the bar)
was turned on gradually and reached full strength at one rotation. We assumed
a single pattern speed for our models, that of the bar, and this was the main
parameter that was varied.
The morphology of the simulated gas distribution was
compared to the morphology observed in the deep $B$-band image of the
galaxy. Since resonance rings generally form in the gas component, the
observed $B$-band morphology is best suited for this purpose. The deep
$B$-band image also lacks the gaps seen in the H$\alpha$ data
(Buta et al. 2001) and is of higher resolution than the existing H I data
(Ryder et al. 1996). We specifically compared the visual size of the inner
ring and outer pseudoring as well as the size and positioning of the
plumes. The size of the nuclear ring is not easy to connect with the ILR of
the bar, therefore the nuclear ring region is not considered in the
visual matching. Regarding the bar itself, dust lanes can be seen along it in
optical images (Figure 1) and there is evidence for it in the
H I intensity map of Ryder et al. (1996).
CO emission was observed by Bajaja et al. (1995) and found to be peaked at
the center of the galaxy and slightly elongated along the bar. Most of the CO
intensity is found within approximately 30$\arcsec$ of the center, or less
than half the bar radius.
In the early phases of gas particle simulations, dust lanes
are often seen but as the evolution proceeds the bar region becomes depleted
of particles. In part, this is because
the relative timescales of the inner and outer parts of the galaxy are not
necessarily realistic. The impact frequency of the gas particles in the
inner ring region is about 25 impacts/particle/bar rotation, whereas in the
outer ring region it is about 2 impacts/particle/bar rotation. The timescale
of evolution is rather arbitrary since it
is determined by the assigned simulation parameters, mainly the radius of the
gas particles, which have no direct physical correspondence to real gas. The
gravitational potential is also static with respect to time which can cause
the rigidly rotating bar to clean some regions of the disk with exaggerated
efficiency. The additional caveat of the simulation code not taking gas
recycling into account leads to a majority of the existing gas particles
collecting at resonance locations.
Overall though, the morphology and size of the bar is a less accurate
pattern speed estimator than the resonance ring and plume features.

\subsection{Results and Discussion}

In Figure 5 (left), we plot the Lindblad precession frequency curves
derived from the simulated rotation curve estimated from the
$H$-band image of the galaxy. These curves show how resonance
locations vary with angular velocity in the linear (epicyclic)
approximation. In this figure (right), we also see that the
simulated outer Lindblad ($\Omega+\kappa/2$), corotation ($\Omega$),
and inner Lindblad ($\Omega-\kappa/2$) frequency
curves appear similar to the same curves derived from the observed
rotation curve data. A significant difference between the simulated
and observed frequency
curves is seen in the inner Lindblad curve. The curve derived from
observational data is seen to reach a maximum of $\Omega-\kappa/2$ $\approx$
5 km s$^{-1}$ arcsec$^{-1}$, whereas the simulated curve has no maximum.
This is due to a subtle difference in the slopes
within a radius of 10$\arcsec$, though the sparsity of observed data points
leaves some ambiguity. The uncertainty of the inner rotation curve, due to
corrections for asymmetric drift and non-circular motions, also propagates to
the Lindblad frequency curves.
A galaxy with a slowly rising rotation curve will have an inner
Lindblad frequency curve that reaches a finite maximum (e.g.,
Yuan \& Kuo 1997). This allows the
possibility for the galaxy to host two inner Lindblad resonances
(ILR), dubbed an inner inner Lindblad resonance (IILR) and outer
inner Lindblad resonance (OILR). If the bar rotates fast enough and
the peak of the $\Omega-\kappa/2$ curve is low, it is also possible
that there is no ILR. A fast rising rotation curve, on
the other hand, will have an inner Lindblad resonance curve with an
infinite maximum and therefore only allow a single ILR. On the other hand,
a secondary bar can also have considerable effect in this region.

In order to estimate $\Omega_p$ for NGC 1433, we would like to find
the value that best matches the simulated and observed morphologies.
For this purpose, we have the inner ring, the plumes, and
the outer pseudoring as features that a good simulation ought to match.
The nuclear ring, however, is not considered in the matching since its
size is much more sensitive to the strength of the perturbation and the method
of the gas treatment.
Good agreement between the size of the inner and outer rings, as
well as the position and orientation of the plumes, is obtained
at 5 bar rotation periods with a pattern speed of 0.89 km s$^{-1}$
arcsec$^{-1}$, or 15.8 km s$^{-1}$ kpc$^{-1}$ with an assumed
distance of 11.6 Mpc (Block et al. 2004), and using the formation
epoch $M/L$ model with bursts from Bell \& de Jong (2001; see Table 1 and
Figure 6). The gravitational potentials derived from the other $M/L$
models
did not cause the simulated morphology of the galaxy to differ
significantly. The plumes found in the simulations are transient
features that are seen to dissipate after about 5 bar rotations.
They begin as complete spiral arms emerging from
a very boxy model inner ring, and evolve to detached segments.
A variation in the pattern speed of
approximately $\pm$5-10\% also prevents the plumes from taking a
shape similar to what is seen in the $B$-band (Figure 7). This
is also true when the bar scale height is reduced.
The inner and
outer ring sizes were only dependent on the bar pattern speed. The
simulated inner ring has a similar size and elongation to the
observed galaxy, but appears more diamond-shaped. The shape of the inner
ring can sensitively depend on the orbit families in this region,
where a rather small change in parameters (e.g. $\Omega_p$ or bar amplitude)
could lead to a large change
in morphology. Since the assumed single bar pattern speed gives both the
inner ring and the structure beyond it, there is no obvious need for a
possible outer, slower spiral mode obtained in other barred spiral
simulations (i.e. Rautiainen \& Salo 1999). It should be noted that
all the simulation timescales are very uncertain. For example, the gas
particle parameters, the rigidity of the gravitational potential, and the
lack of gas recycling in the system all affect the dynamical timescale of
the simulation.

Examining the morphology of the galaxy, we find that the
observed inner ring corresponds to the location of the simulated
inner 4:1 resonance ($\Omega_p=\Omega-\kappa/4$) and the outer pseudoring
lies between the corotation resonance and
outer 4:1 resonance ($\Omega_p=\Omega+\kappa/4$) (Figure 8).
The plumes and the outer pseudoring lie between the inner and outer
($\Omega+\kappa/4$) 4:1 resonances, similar to what was determined for
ESO 566-24 (Rautiainen et al. 2004). An outer ring was also shown to
occasionally form
within the outer 4:1 resonance, instead of near the OLR, in an N-body
model produced by Rautiainen \& Salo (2000). This suggests that some
outer pseudorings that show the R$_1^{\prime}$ morphology (Buta \& Crocker
1991) are not actually OLR features. This was already hinted at
observationally by ESO 566$-$24, where two of the four outer arms
were noted to form an R$_1^{\prime}$ morphology by Buta et al. (1998).
In this galaxy, a clear detached outer ring that encircles the $m=4$
pattern could trace the actual location of the OLR.

These results may be compared with the precession frequency analysis
made by Buta et al. (2001, their Fig. 14), who used preliminary numerical
simulations from Buta \& Combes (2000) to set the pattern speed at
1.3 km s$^{-1}$ arcsec$^{-1}$. These simulations did not reproduce the
plumes of NGC 1433, and the derived pattern speed placed the outer
Lindblad resonance closer to the radius of the outer pseudoring than
do our present simulations. In Table 2, we summarize the resonance
radii derived by both Buta et al. (2001) and our best model. Our model
predicts that the OLR is at a radius about 1.4 times larger than the observed
major axis radius of the outer pseudoring. The radial
distribution of the azimuthally averaged H I surface density published by
Ryder et al. (1996, their Fig. 6) shows that there is very little gas in
the OLR region predicted by our new simulations. This may explain why no
features associated with this resonance are observed in NGC 1433.

An obvious difference between the simulated and observed
morphologies is the size of the nuclear ring. The nuclear ring seen
in the simulations is much larger than what is observed in the
actual galaxy. According to Buta \& Combes (1996), nuclear rings
tend to form near the IILR if it exists. By plotting our value of
$\Omega_p$ over the Lindblad precession frequency curves in Figure
5, we are able to predict where resonances occur by determining the
radius at which the $\Omega_p$ line crosses a corresponding
frequency curve. For the observed inner Lindblad frequency curve,
the smallest intersection radius is near 3$\arcsec$. In
Table 1 of Buta et al. (2001), the semimajor axis radius of the
nuclear ring is given as 0.5 kpc at a distance of 11.6 Mpc, or
8.9$\arcsec$, which is about three times the estimated linear resonance
radius. In the case
of the simulation, the fast rising rotation curve forces the nuclear
ring to form near the single ILR which corresponds to the OILR derived
from the observed rotation curve.
The presence of a secondary
bar with a different pattern speed is a further complication. The
observed nuclear ring can be an example of loops which were introduced by
Maciejewski \& Sparke (2000) in orbit analysis in double barred
potentials and simulated by Rautiainen \& Salo (2002). In principle, a
stronger bar perturbation, due to the bar being less vertically extended
for example, would shift the nuclear ring inward from the linear ILR
distance (see e.g. Salo et al. 1999). However, decreasing the
scale height of the bar did not have a significant effect on the
size of the nuclear ring. In contrast, Regan \& Teuben (2003)
have stated that the ILR does not exist for a galaxy with a strong bar,
like NGC 1433, since there is no resonance with the rotating potential
at that radius. Instead, they showed through their hydrodynamical models
that nuclear rings result from the
interaction of gas on $x_1$-like and $x_2$-like streamlines.
This suite of uncertainties in the central region of NGC 1433 indicates
that the observed nuclear ring does not serve as an accurate
indicator of $\Omega_p$.

In an effort to examine the expected deviations from the linear epicyclic
approximation caused by a strong bar, we have produced models of different
bar amplitudes (e.g. the non-axisymmetric Fourier components of the potential
were multiplied by a factor while the axisymmetric component was kept
intact). In Figure 9, we compare the observed gas morphology
to models at 5 bar rotations with $\Omega_p$ equal to 0.89 km s$^{-1}$
arcsec$^{-1}$ and with bar amplitudes varied by a factor of 0.25 to 1.25.
As mentioned earlier, these models also serve to describe the possible effects
of uncertainties in factors like bar height or the contribution from the halo.
The results are similar to those found when the bar amplitude of IC 4214 was
varied by Salo et al. (1999).
As the amplitude is increased from a factor of 0.25 to 1.00, the inner ring
becomes well defined and increases in ellipticity. The ring nearly vanishes
when
the amplitude is increased by a factor of 1.25. The resultant morphology from
varying the amplitude indicates that the shape
of the observed inner ring may be due to a somewhat weaker perturbation
than assumed in the nominal model. Also, as the bar amplitude of the model
is increased, the size of the nuclear ring decreases.

In Figure 10, we compare the observed and simulated kinematics of the galaxy
by examining the zero-velocity contour, which is sensitive to bar-induced
radial velocities.
We compare the
H$\alpha$ Fabry-Perot velocity field found in Buta et al. (2001) with
the simulated gas velocity fields derived from various bar amplitudes.
Agreements between the observed and simulated kinematics of the
galaxy support our estimate of $\Omega_p$ with a bar amplitude factor
close to 1.
Although the observed
velocity field is not as rich with data as the simulated velocity
field, the zero-velocity contours show similar characteristics
within 100$\arcsec$ from the galaxy center. A comparison
of the observed and modeled major axis velocity profiles (Figure 11),
sensitive to bar induced tangential velocity increments, reveals a
similar behavior between approximately 50$\arcsec$ and 80$\arcsec$.
A deviation is seen between the azimuthally averaged total velocity
contribution
calculated from the mass model and the major axis velocity profile
near 30$\arcsec$ due to the large nuclear ring in the model.
Particles in this
ring, near the major axis, are close to their apocenter and are therefore
moving slower than their average orbital velocity. At larger
radii, the modeled major axis velocity profile is approximately
15 km s$^{-1}$ less than
the azimuthally averaged total velocity
contribution used in the simulations. In principle, the match in this region
could be easily improved by assuming a somewhat larger halo contribution. This
would also reduce the relative bar perturbation.

Also, the asymmetry seen in the simulated radial
velocity of the gas found in the inner ring of the galaxy is similar to
what is observed (Figure 12).
The upper right panel of this figure shows the velocity-position angle
diagram for points confined between the two ellipses shown in the upper left
panel. The observed velocity-position angle diagram shows a subtle asymmetry
that was also detected by Buta (1986) and interpreted in terms of ring
material moving along an oval bar orbit. The lower right panel of
Figure 12 shows the same diagram for the simulated inner ring, showing
a similar asymmetry. The main difference is that the simulated curve
is steeper near position angles of 100$^{\circ}$ and 280$^{\circ}$ owing to
the more
pointy oval shape of the simulated inner ring. By fitting a simple
orbit model to the observed velocity-position angle diagram, Buta (1986)
estimated a pattern speed of 1.55 $\pm$ 0.4 km s$^{-1}$ arcsec$^{-1}$. If
the bar were exactly perpendicular or parallel to the line of nodes, this
asymmetry would not be seen, as in the case of NGC 3351 (Buta 1988).

Because of the versatility of the modeling process, we are able to
track a number of individual gas particles for the duration of a
simulation. In Figure 13 we show the paths of selected gas particles
in the rotating frame of the simulated galaxy from 1.0 (when the bar
has reached full strength) to 5.0 bar
rotations where $\Omega_p$ = 0.89 km s$^{-1}$ arcsec$^{-1}$. We
find that the nuclear ring is partially made up of gas particles that
originate from near the final location of the inner ring. The
inner ring includes some particles drawn from outside the final
inner ring location. The transient plumes seen in the simulations
appear to be due to both gas trapped in the outer banana orbits and traveling
from theses orbits inward toward the inner ring.

Determining a value for $\Omega_p$ also allows us to estimate
$\cal{R}$ $\equiv R_{CR}/R_{B}$ where $R_{CR}$ is the corotation
radius of the galaxy and $R_{B}$ is the bar semimajor axis length.
If this distance-independent ratio is between 1.0 and 1.4, a bar is
said to be ``fast,'' while if greater than 1.4, a bar is said to be
``slow.'' Contopoulos (1980) concluded that self-consistent bars cannot
exist when $\cal{R} <$ 1.0, although Zhang \& Buta (2007)
have argued to the contrary. Debattista \& Sellwood (2000) contend
that fast bars exist in halos with a low central concentration since
the bar rotation rate would rapidly decrease due to dynamical
friction with the halo. Certain galaxy models from Athanassoula
(2003) also show this correlation between fast bars and halos of low
central concentration. The value of $R_{CR}$ is found by determining
the radius at which $\Omega_p=\Omega$ in the simulation. For our
$\Omega_p$ value of 0.89 $\pm$ 0.06 km s$^{-1}$ arcsec$^{-1}$, the
average $R_{CR}$ value is
162.5 $\pm$ 11.9$\arcsec$ (see Figure 5).

The deprojected value of $R_{B}$ was estimated to be 82$^{\arcsec}$
(4.6 kpc) by Buta et al. (2001). With our average value of $R_{CR}$, this
would imply an upper limit of $\cal{R}=$ 2.0. This is because
$R_{B}=$ 82$^{\arcsec}$ refers to the apparent sharp
ends of the bar, and may not be the best estimate of the actual bar radius.
Based on a bar-spiral Fourier separation analysis, Figure 1 of
Block et al. (2004) suggests that the bar extends to a maximum radius
of 115 $\pm$ 5$\arcsec$. This value implies that the lower limit of
$\cal{R}$ is 1.4 $\pm$ 0.1. An average of these $R_{B}$ values
(98.5$\arcsec$) yields $\cal{R}=$ 1.7 $\pm$ 0.3, with the error largely due to
the ambiguity in the bar length.
In any case, NGC 1433 may be in the ``slow bar" domain of Debattista
\& Sellwood (2000). Most previous direct measurements of bar pattern speeds
in early-type galaxies have not found slow bars (e.g. Corsini et al. 2007; 
Debattista \& Williams 2004; Aguerri et al. 2003; Corsini et al. 2003).

\section{Conclusions}

Using near-IR images to trace the stellar disk potential, we
have located resonances in NGC 1433, an exceptional barred spiral
galaxy having strong morphological features.  The numerical
simulation method of Salo et al. (1999; see also Rautiainen et al.
2005) was used to interpret the structure of the
galaxy by matching a cloud-particle model to the observed
morphology. The results are consistent
with previous studies such as Schwarz (1984), Byrd et al. (1994),
Rautiainen \& Salo (2000), and Rautiainen, Salo, \& Buta (2004).
Our best simulation model provides
an interpretation of NGC 1433 which places corotation at 1.7 times
the average estimated bar radius of about 100$^{\arcsec}$. This model
also places the inner 4:1 resonance near the cuspy ends of
the highly elongated inner ring and just outside the ends of the bar.

The plumes and the two outer arms forming the R$_1^{\prime}$ outer
pseudoring appear confined completely between the inner and outer
4:1 resonances, very similar to what Rautiainen, Salo, \& Buta (2004)
found for the symmetric $m=4$ barred spiral ESO 566$-$24. The
plumes, which are like detached spiral segments,
appear to be evolved versions of the two
extra ``side arms" of ESO 566$-$24 that wrap around the bar ends.

The predicted location of the OLR in NGC 1433 is well outside
the outer pseudoring in a region where there is little H I gas
(Ryder et al. 1996).
Thus, NGC 1433 presents an example of an ``OLR subclass" outer
pseudoring (Buta \& Crocker 1991; Buta, Corwin, \& Odewahn 2007)
that is {\it not} associated with the OLR.


P. Treuthardt and R. Buta acknowledge the support of NSF Grants AST-0205143
and AST-0507140 to the University of Alabama. H. Salo and P. Rautiainen
acknowledge the support of the Academy of Finland.
P. Treuthardt acknowledges the support of the
Academy of Finland during two summer visits to Oulu in 2002 and 2003.

\clearpage


\clearpage

\input{tab1.tex}

\clearpage

\input{tab2.tex}

\clearpage

\begin{figure}
\figurenum{1}
\includegraphics[scale=0.21,viewport=54 14 1029 950]{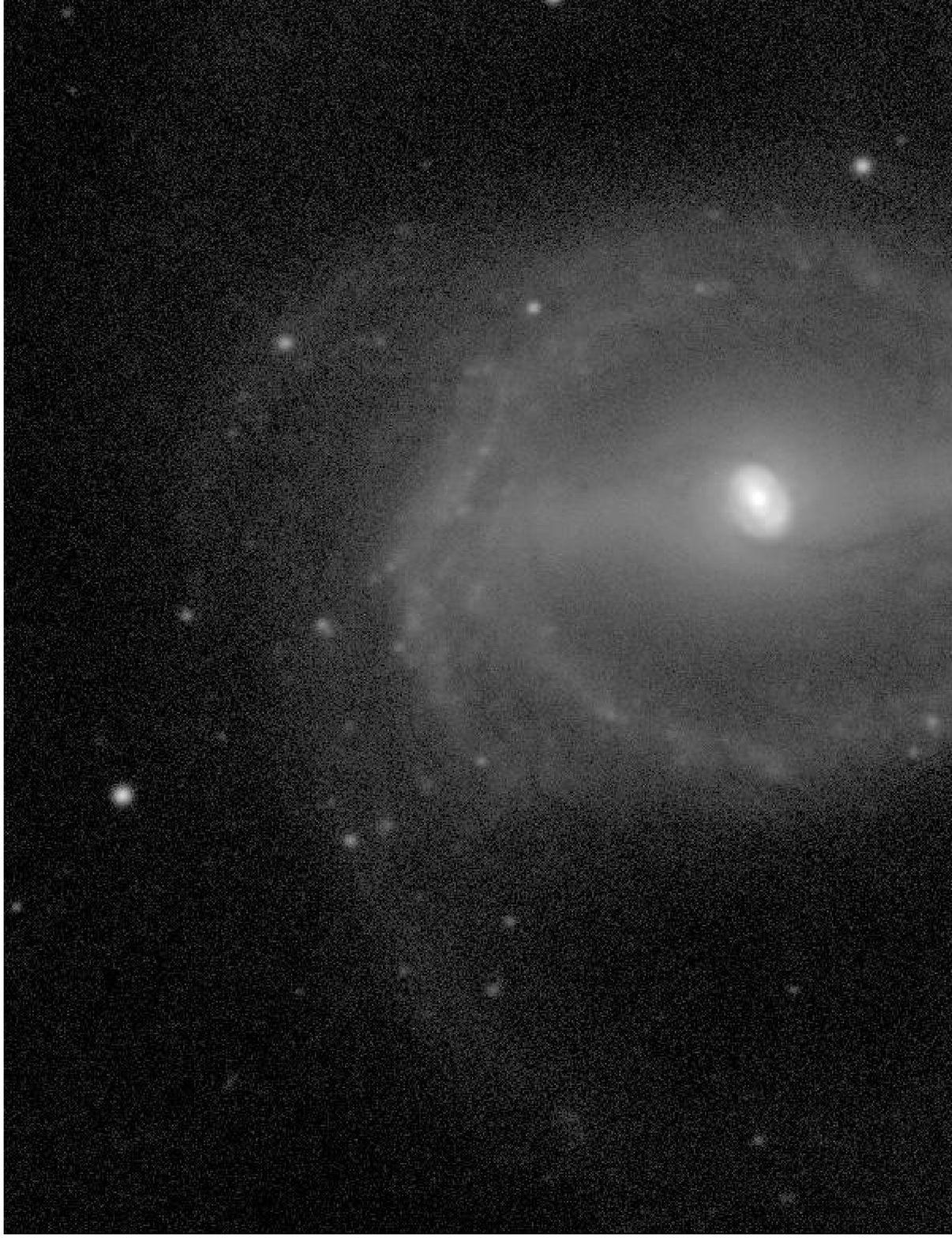}
\includegraphics[scale=0.21,viewport=14 14 1029 948]{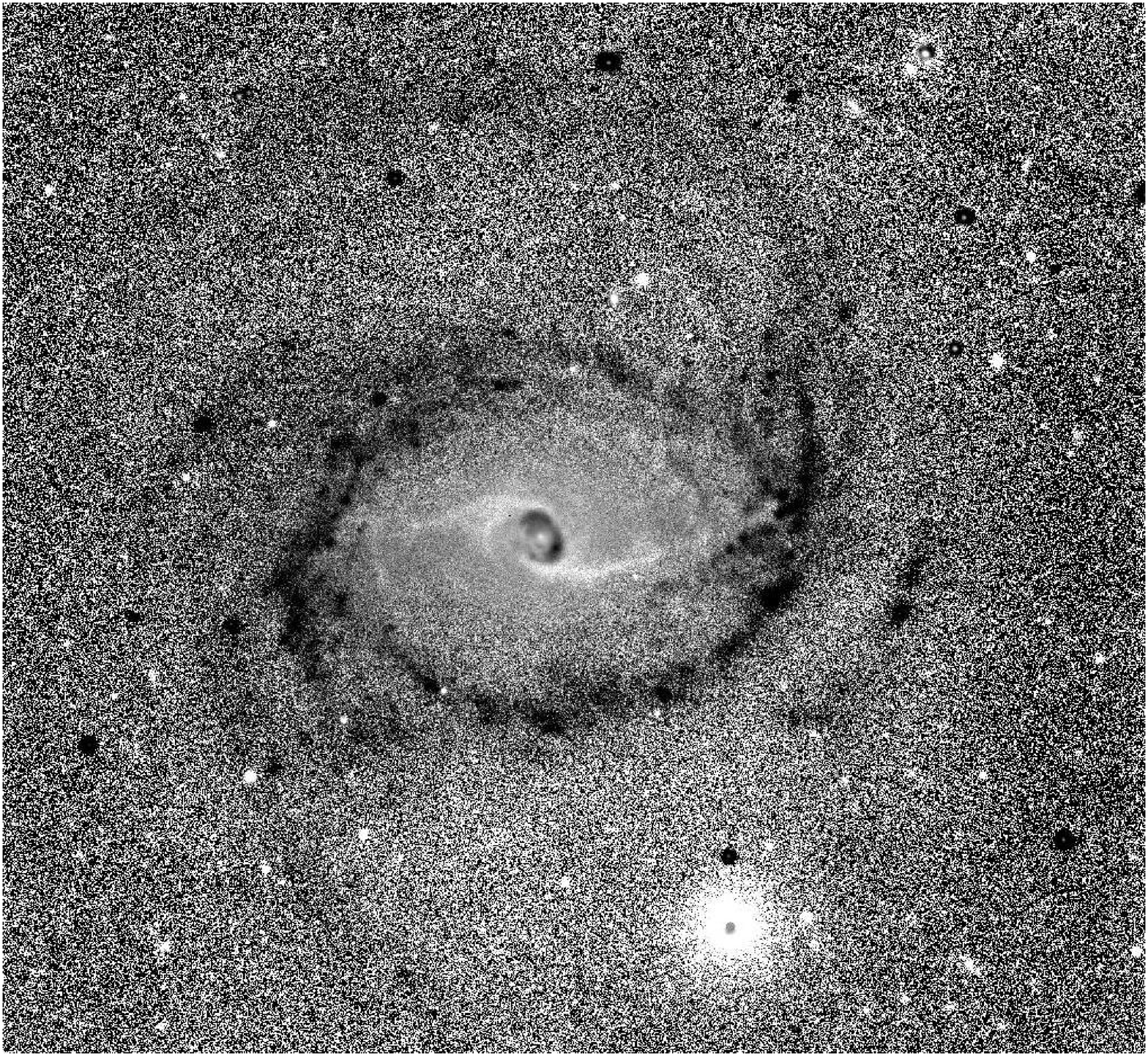}
\includegraphics[scale=0.42,viewport=28 42 538 679]{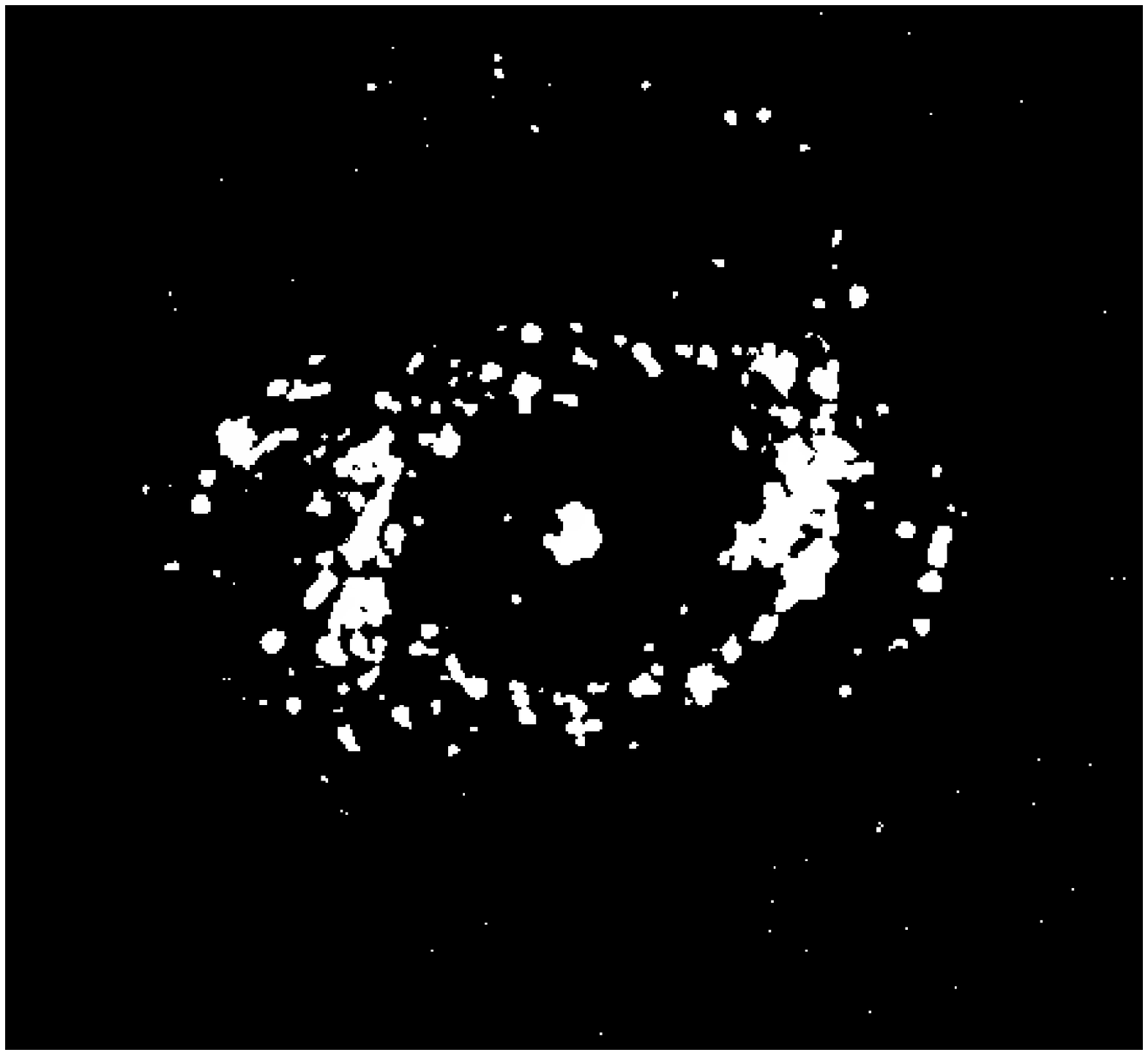}
\includegraphics[scale=0.42,viewport=-78 42 538 679]{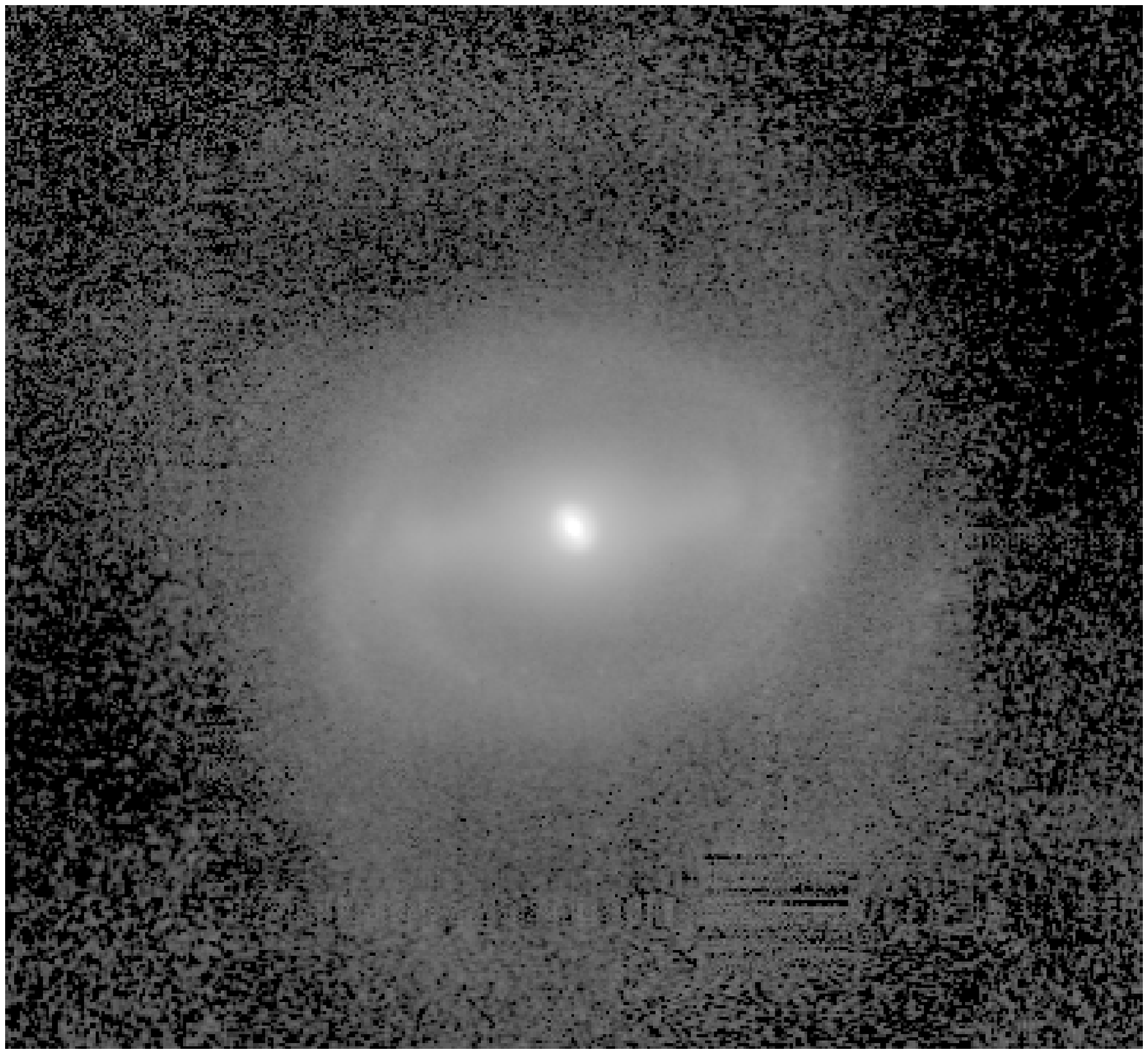}
\label{sample}
\caption{$B$-band (upper left), $B-I$ color index (upper right), H$\alpha$
Fabry-Perot (lower left), and $H$-band (lower right) images displaying the
morphology of NGC 1433. In the color index image, the dark areas represent
bluer regions, while the white areas represent redder regions. All images are
7.4$\arcmin$ by 6.8$\arcmin$. The $B$ and $B-I$ images are taken from Buta
et al. 2007 while the H$\alpha$ and $H$-band images are taken from Buta
et al. 2001.}
\end{figure}

\clearpage

\begin{figure}
\figurenum{2}
\includegraphics[angle=90,height=125mm]{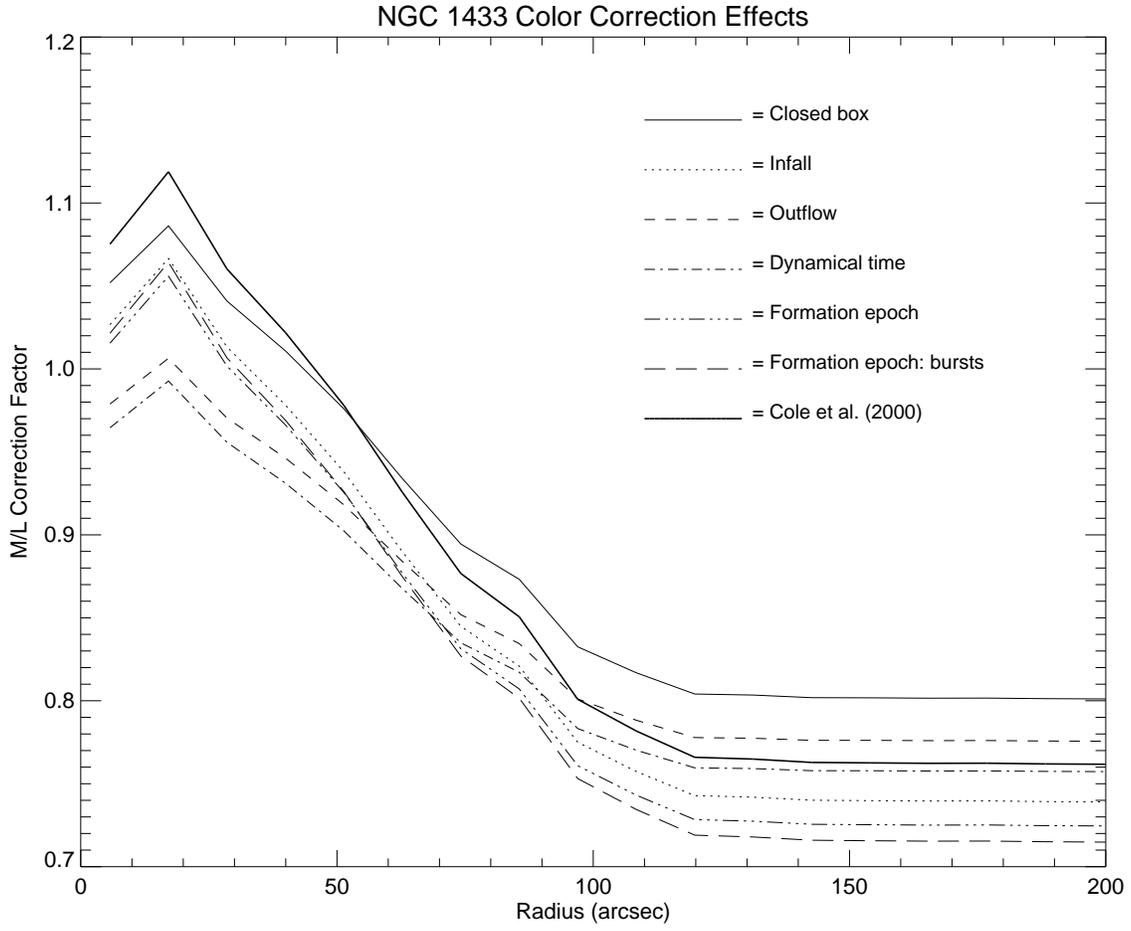}
\label{sample}
\caption{A plot of the radial M/L profiles derived
from each of the seven models described in Bell \& de Jong 2001.}
\end{figure}

\clearpage

\begin{figure}
\figurenum{3}
\includegraphics[angle=90,height=125mm]{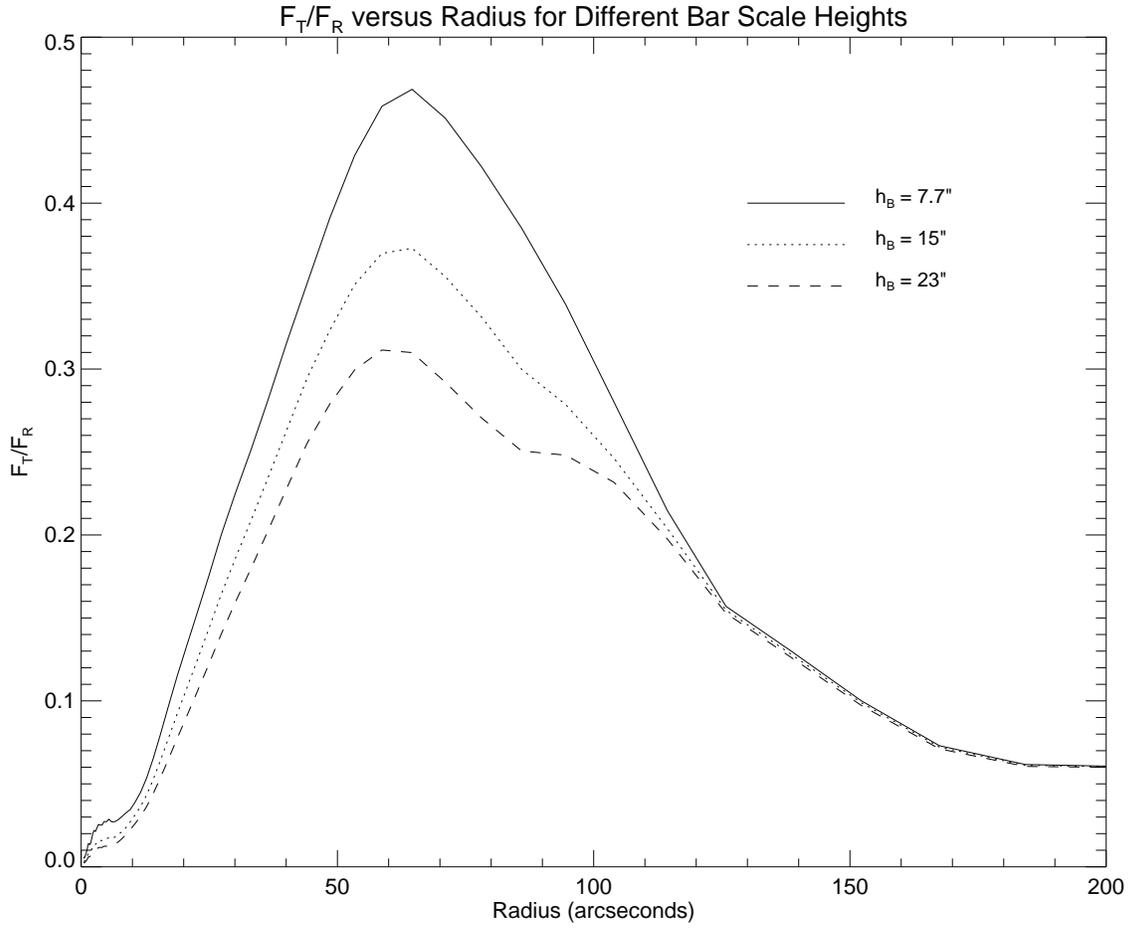}
\label{sample}
\caption{Plots of the tangential force amplitude ($F_{T}$) normalized
by the axisymmetric disk force ($F_{R}$) for different bar scale heights
($h_{B}$). The bar scale heights range from one to three times the disk scale
height of 7.7$\arcsec$.}
\end{figure}

\clearpage

\begin{figure}
\figurenum{4}
\includegraphics[angle=90,height=125mm]{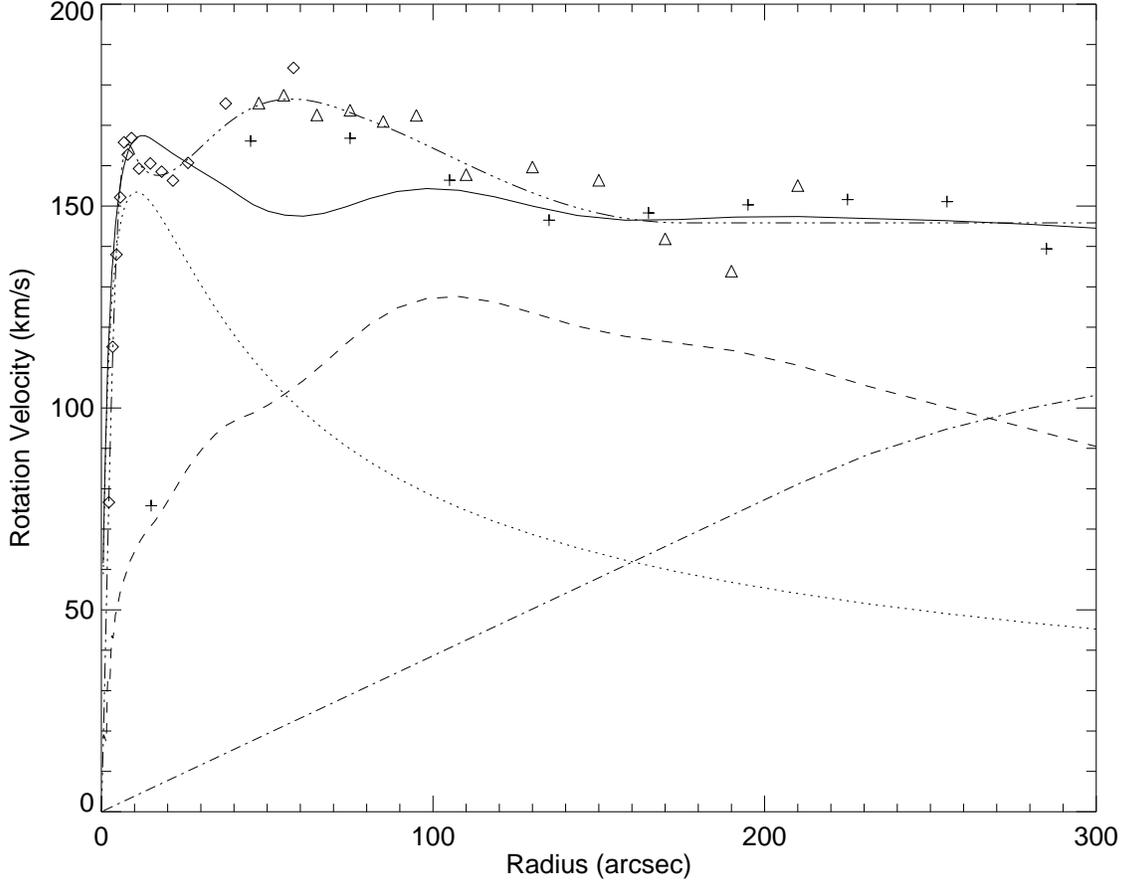}
\label{sample}
\caption{A plot of both the observed composite rotation curve
(triple-dot-dashed line) and the simulation rotation curve (solid
line) for NGC 1433. The composite rotation curve is from Buta et
al. (2001), based on a polynomial fit to H I, H$\alpha$, and Mg I b
velocity data (plus, triangle, and diamond symbols, respectively),
corrected for an inclination of 33$^{\circ}$. The points have typical
statistical uncertainties of $\pm$10-20 km s$^{-1}$, due to non-circular
motions and an uncertain correction for asymmetric drift within 60$\arcsec$.
Lines are also shown that indicate the azimuthally averaged velocity
contributions
of the bulge (dotted), disk (dashed), and halo (dot-dashed) used in the
simulations. These lines were calculated from the mass model based on the
formation epoch
with bursts (Bell \& de Jong 2001) M/L corrected NIR image of the galaxy,
an assumed disk vertical scale height of 7.7$\arcsec$, and a distance of
11.6 Mpc.
The Persic halo component was modeled as an isothermal sphere with a core
radius of 213$\arcsec$ and an asymptotic velocity at infinity of 127 km
s$^{-1}$.
The solid line is the sum of the component contributions or the total simulated
rotation curve.}
\end{figure}


\clearpage

\begin{figure}
\figurenum{5}
\includegraphics[angle=90,height=63mm]{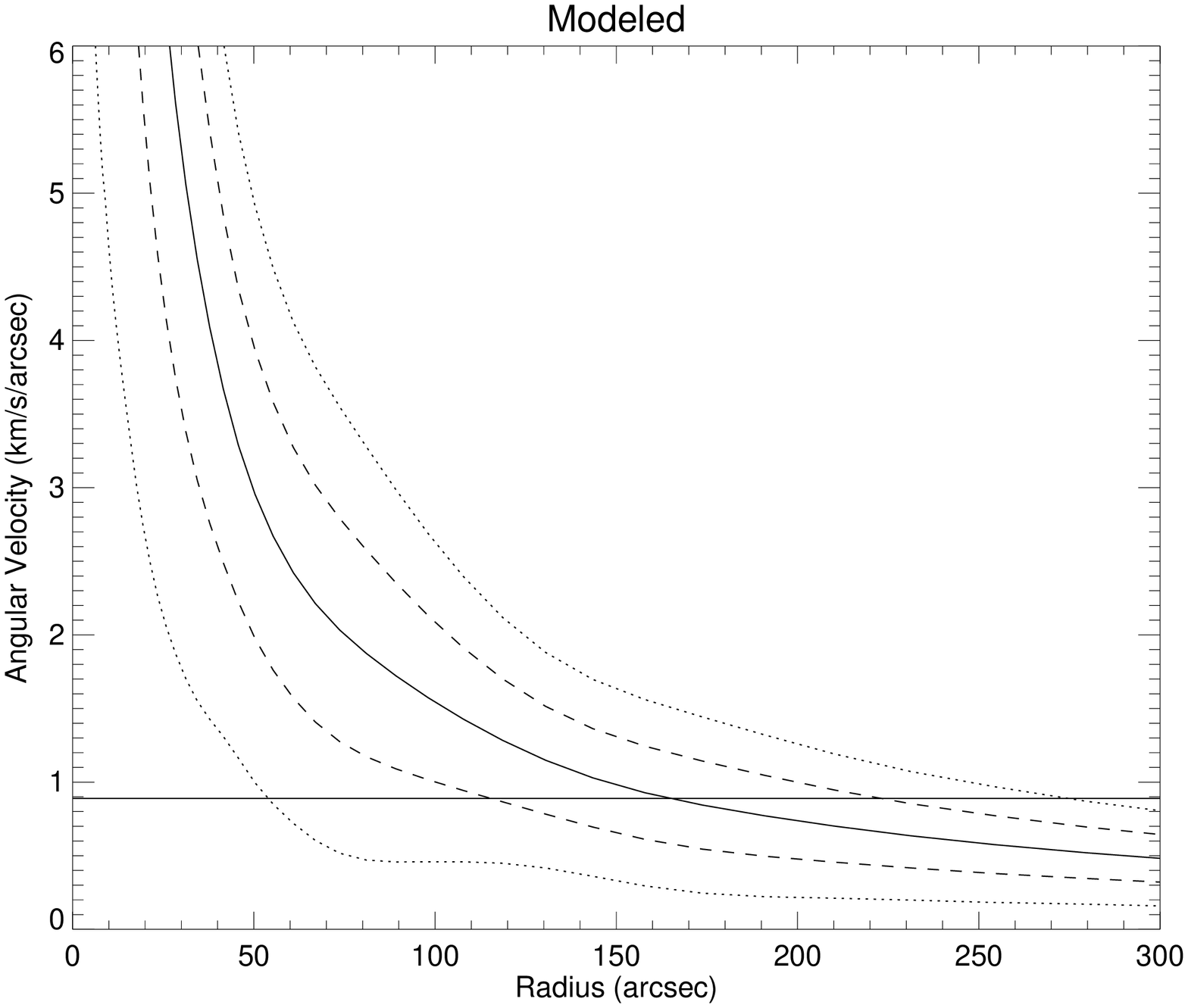}
\includegraphics[angle=90,height=63mm]{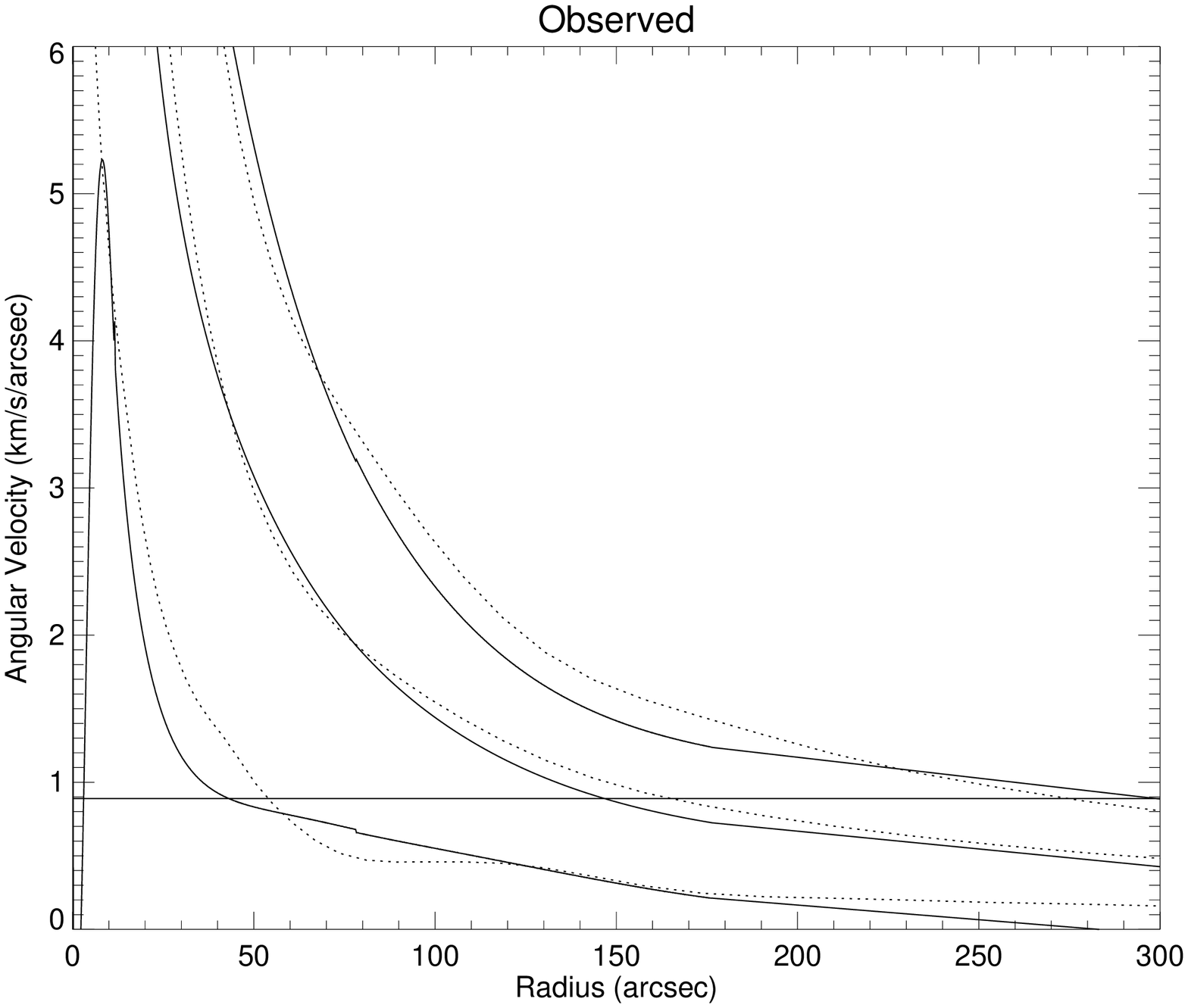}
\caption{Lindblad precession frequency curves for NGC 1433. The horizontal
line in both plots corresponds to the inclination corrected $\Omega_p$ of 0.89
km s$^{-1}$ arcsec$^{-1}$ or 15.8 km s$^{-1}$ kpc$^{-1}$ with a galaxy
distance of 11.6 Mpc. The left plot shows the frequency curves derived from
the simulated rotation curve shown in Figure 2.
The horizontal line intersects curves corresponding to, from smaller to
larger radii, $\Omega-\kappa/2$, $\Omega-\kappa/4$, $\Omega$,
$\Omega+\kappa/4$, and $\Omega+\kappa/2$. $\Omega$ is the circular angular
velocity and $\kappa$ is the epicyclic frequency. The corotation radius is
where $\Omega=\Omega_p$. The right plot displays the simulated
$\Omega-\kappa/2$, $\Omega$, and $\Omega+\kappa/2$ curves (dotted) with the
corresponding curves derived from the observed composite rotation curve from
Buta et al. (2001) overlayed (solid).
}
\end{figure}

\clearpage

\begin{figure}
\figurenum{6}
\includegraphics[angle=0,scale=1.0]{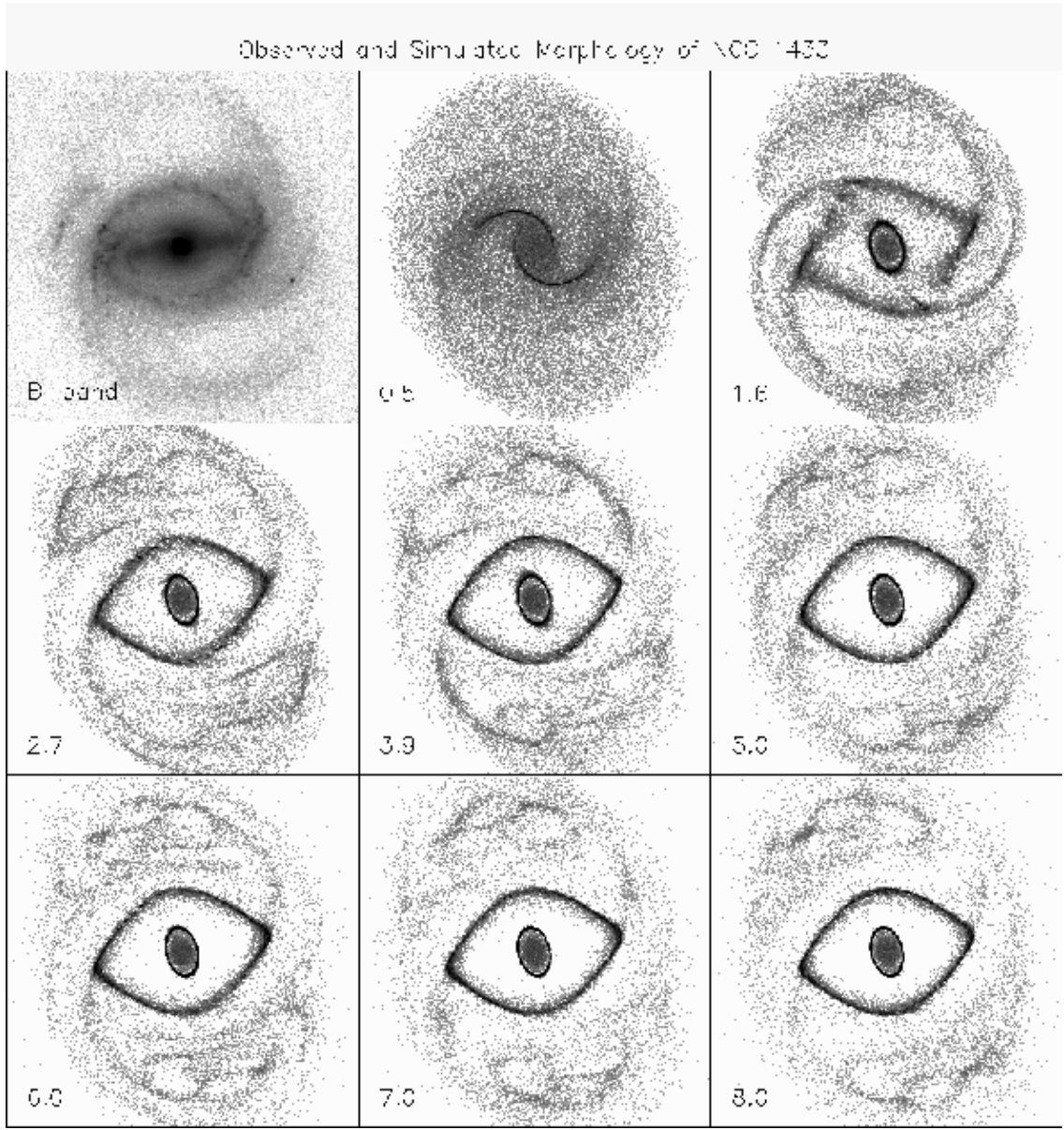}
\label{sample}
\caption{A $B$-band image of NGC 1433 (upper left panel) showing
the (R$^{\prime}$$_1$)SB(rs)ab morphology. The remaining panels are
gas simulations showing the sky-plane morphology at 0.5, 1.6, 2.8, 3.9, 5.0,
6.0, 7.0, and 8.0 bar
rotations
using a pattern
speed of 0.89 km s$^{-1}$ arcsec$^{-1}$. The simulated bar has a scale height
3 times larger than that of the disk. After a few bar rotations, the simulated
and observed morphologies appear similar. Note the ``plumes,'' or detached
arm segments, to the upper left and lower right of the inner ring in the
observed and simulated morphologies. All panels are 400 by 400 arcseconds.}
\end{figure}

\clearpage

\begin{figure}
\figurenum{7}
\includegraphics[angle=0,height=120mm]{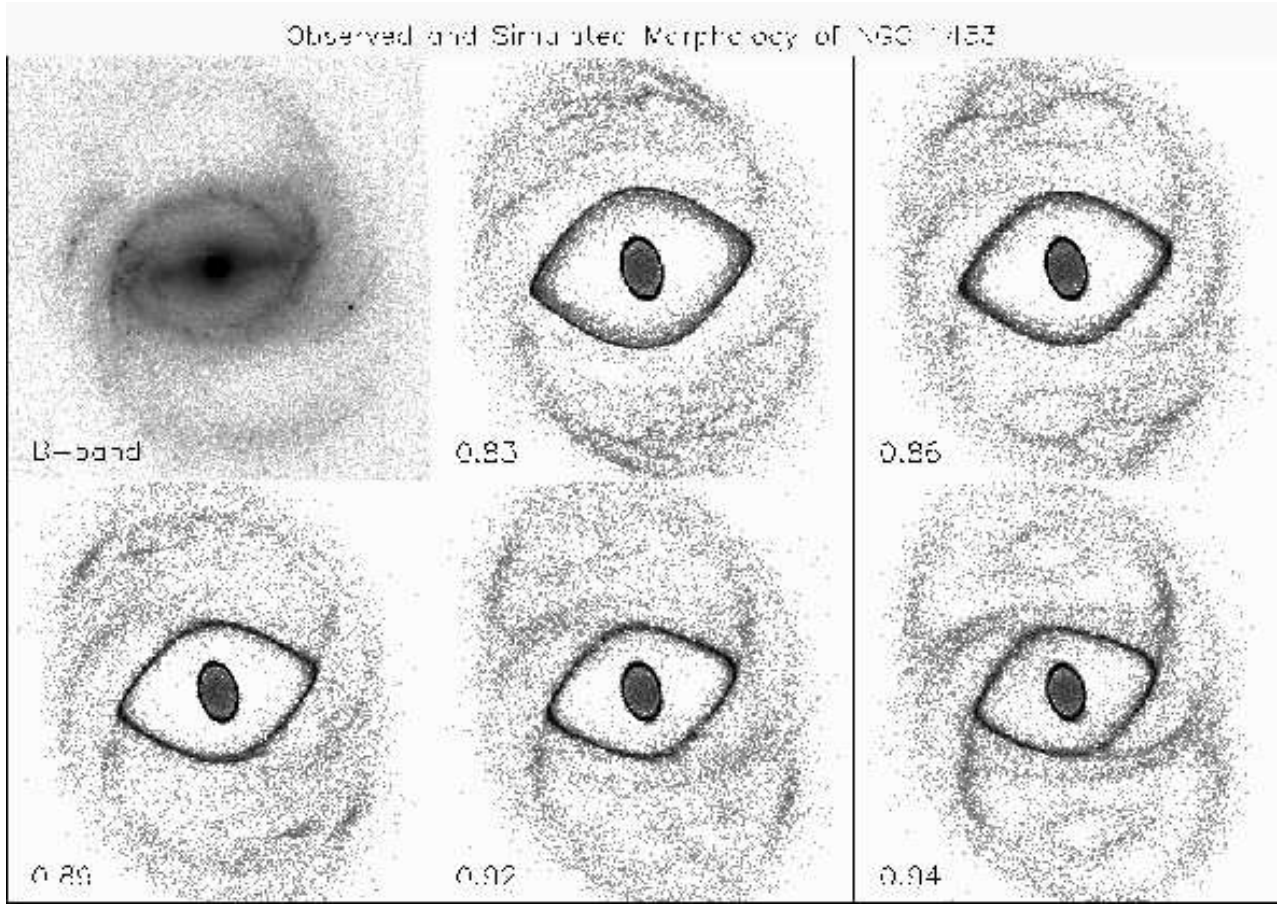}
\label{sample}
\caption{A $B$-band image of NGC 1433 (upper left panel) showing
the (R$^{\prime}$$_1$)SB(rs)ab morphology. The remaining panels are
gas simulations showing the morphology at 5.0 bar rotations with pattern
speeds of 0.83, 0.86, 0.89, 0.92, and 0.94 km s$^{-1}$ arcsec$^{-1}$. The
shape and position of the simulated plumes most closely resembles the $B$-band
morphology with a pattern speed of 0.89 km s$^{-1}$ arcsec$^{-1}$ is used. Also
note the decrease in size of the inner ring with the increase in pattern speed.
All panels are 400 by 400 arcseconds.}
\end{figure}

\clearpage

\begin{figure}
\figurenum{8}
\includegraphics[angle=90,height=125mm]{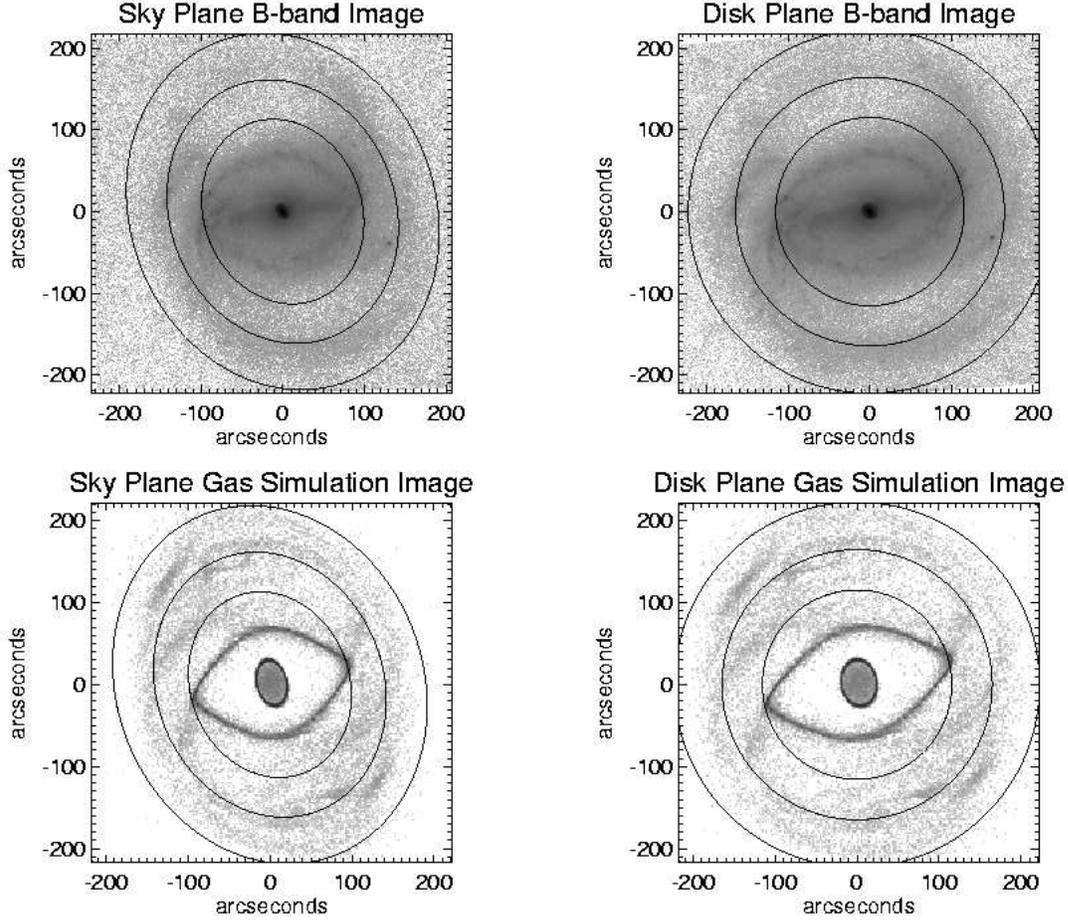}
\label{sample}
\caption{$B$-band disk- (upper left) and sky-plane (upper right) images
of NGC 1433 along with disk- (lower left) and sky-plane (lower right) gas
particle simulations shown at 5.0 bar rotations with a pattern speed of
0.89 km s$^{-1}$ arcsec$^{-1}$. The overlayed ellipses are resonance
locations derived from the shown gas particle simulations. The inner-most,
middle, and outer-most ellipses
correspond to the $\Omega-\kappa/4$, corotation, and $\Omega+\kappa/4$
resonances, respectively.}
\end{figure}

\clearpage

\begin{figure}
\figurenum{9}
\includegraphics[angle=0,height=120mm]{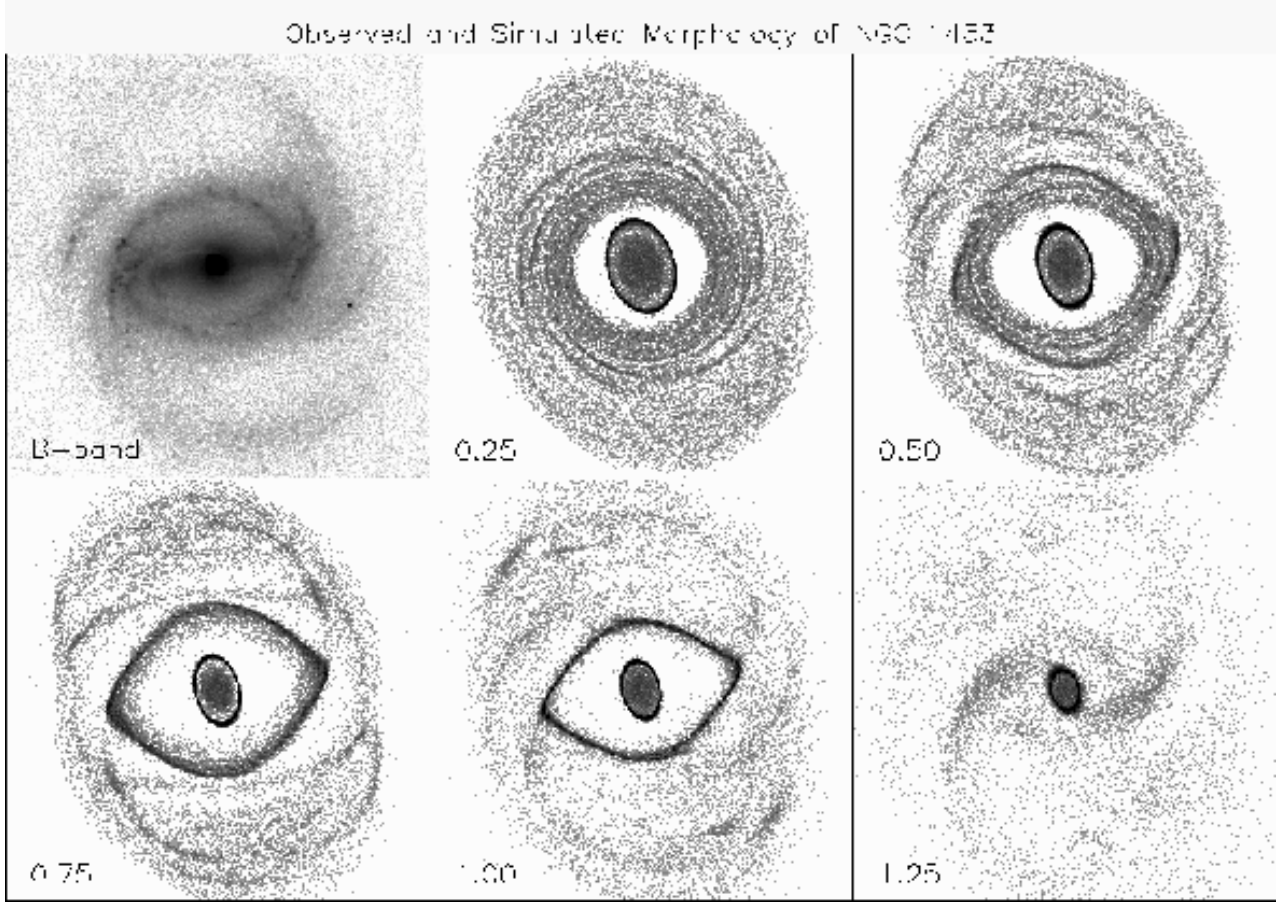}
\label{sample}
\caption{A $B$-band image of NGC 1433 (upper left panel) showing
the (R$^{\prime}$$_1$)SB(rs)ab morphology. The remaining panels are
gas simulations showing the morphology at 5.0 bar rotations with a pattern
speed of 0.89 km s$^{-1}$ arcsec$^{-1}$ and bar amplitudes varying from 0.25 to
1.25. All panels are 400 by 400 arcseconds.}
\end{figure}

\clearpage

\begin{figure}
\figurenum{10}
\includegraphics[angle=90,scale=0.73]{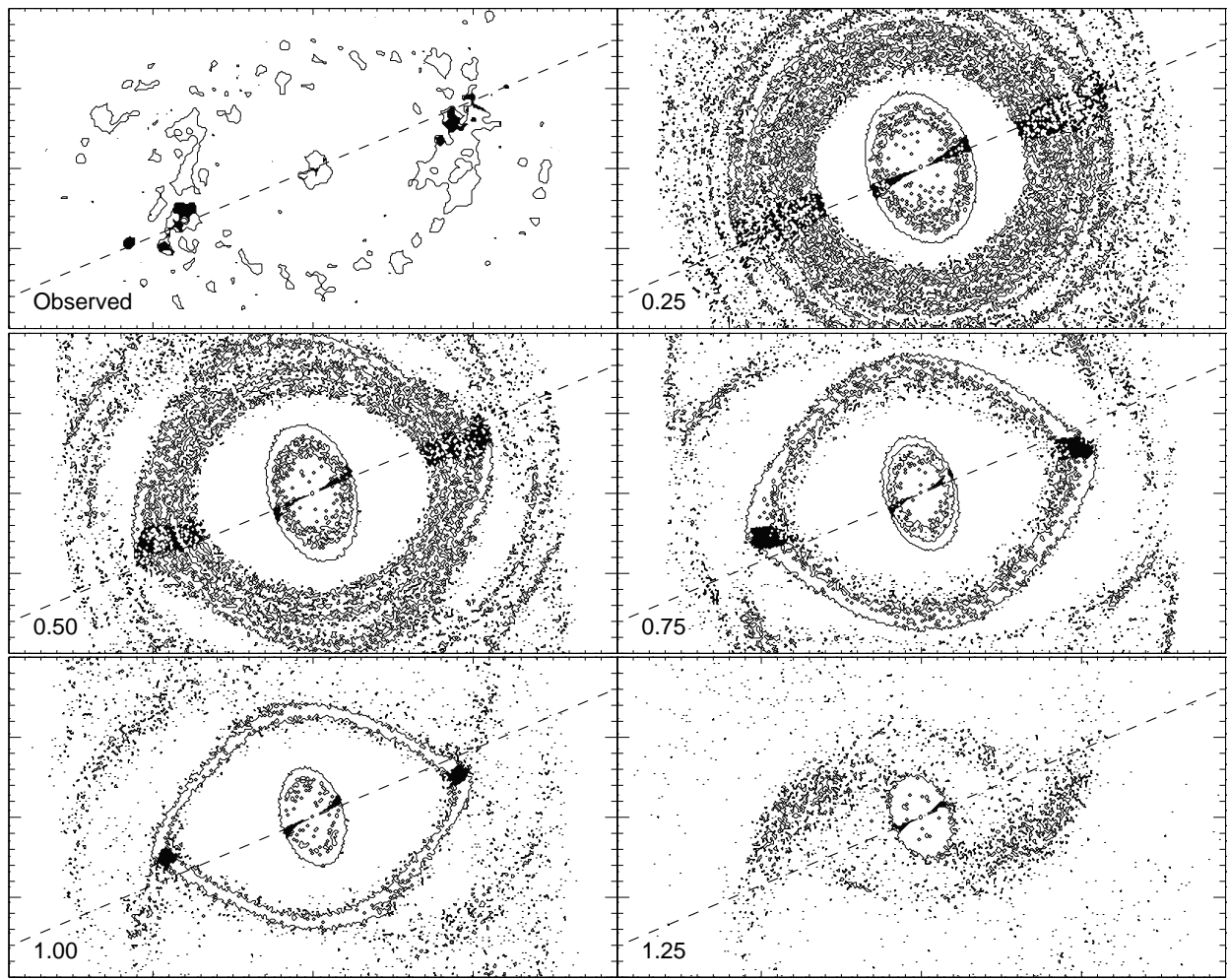}
\label{sample}
\caption{Zero-velocity regions (filled contours) from the observed and
simulated data derived from velocities within $\pm$10 km s$^{-1}$ of the
systemic velocity.
The upper left frame shows the H$\alpha$ velocity field of NGC 1433 from
Buta et al. 2001 while the remaining frames show the gas particle distribution
at 5.0 bar rotations with an $\Omega_p$ of 0.89 km s$^{-1}$ arcsec$^{-1}$ and bar
amplitude factors ranging from 0.25 to 1.25. The
dashed line in the frames coincide with the galaxy's minor axis. All frames are
380 by 200 arcseconds.}
\end{figure}

\clearpage

\begin{figure}
\figurenum{11}
\includegraphics[angle=90,height=125mm]{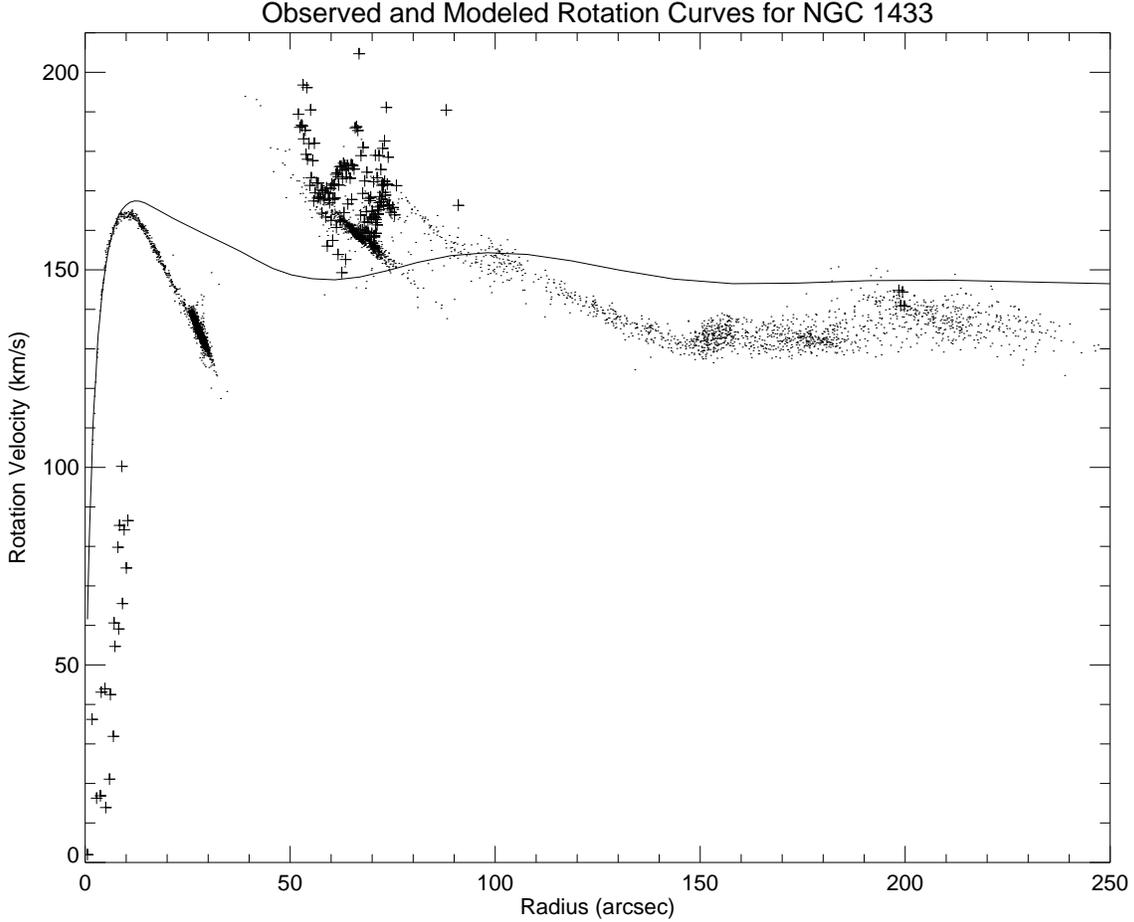}
\label{sample}
\caption{A plot of the observed (plus symbols) and modeled
(points) major axis velocity profiles for NGC 1433, as well as the
azimuthally averaged total velocity contribution calculated from
the mass model (line) discussed in Figure 4. The major axis velocity
profiles are derived
from the observed H$\alpha$ velocity field (from Buta et al. 2001) and the
modeled gas particle
distribution at 5 bar rotations with an $\Omega_p$ of 0.89 km s$^{-1}$
arcsec$^{-1}$. The velocity profile data was extracted from within 5$^{\circ}$
of the major axis position angle of 21$^{\circ}$ and corrected for an
inclination of 33$^{\circ}$. The deviation between the model and the observed
gas velocities for radii less than 20$\arcsec$ is due in part to noncircular
motions, manifested as a 9$^{\circ}$ counterclockwise
offset between the kinematic major axis
of the inner velocity field in Fig. 6 of Buta et al. (2001)
and the overall position angle. The deviation is also due to the model,
which we have shown does not describe the inner regions
reliably owing to our assumption of a single pattern speed
and other model details.}



\end{figure}

\clearpage

\begin{figure}
\figurenum{12}
\includegraphics[angle=90,height=125mm]{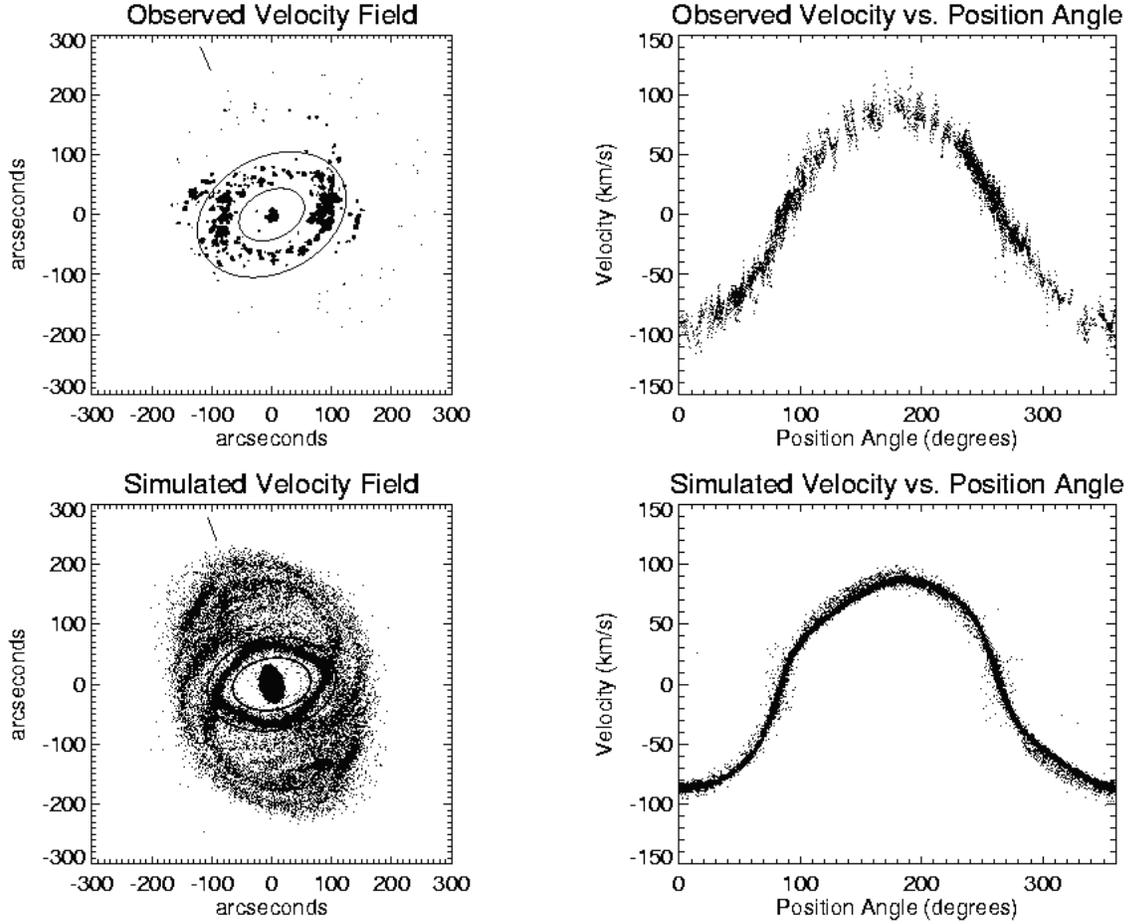}
\label{sample}
\caption{Line-of-sight velocities along the inner ring of NGC 1433. The
left-most
plots show the annuli where the inner ring velocity data was extracted from
the observed H$\alpha$ velocity field from Buta et al. 2001 (upper) and the
simulated gas particle distribution at 5.0 bar rotations with an $\Omega_p$
of 0.89 km s$^{-1}$ arcsec$^{-1}$. The angled line near the top of the frames
coincides with the line-of-nodes of the galaxy. The right-most plots show the
line-of-sight velocity versus the position angle relative to the
line-of-nodes along the ring from the observed
(upper) and simulated (lower) data. There is a clear asymmetry in both the
observed and simulated radial velocity data along the inner ring.}
\end{figure}

\clearpage

\begin{figure}
\figurenum{13}
\includegraphics[angle=90,height=125mm]{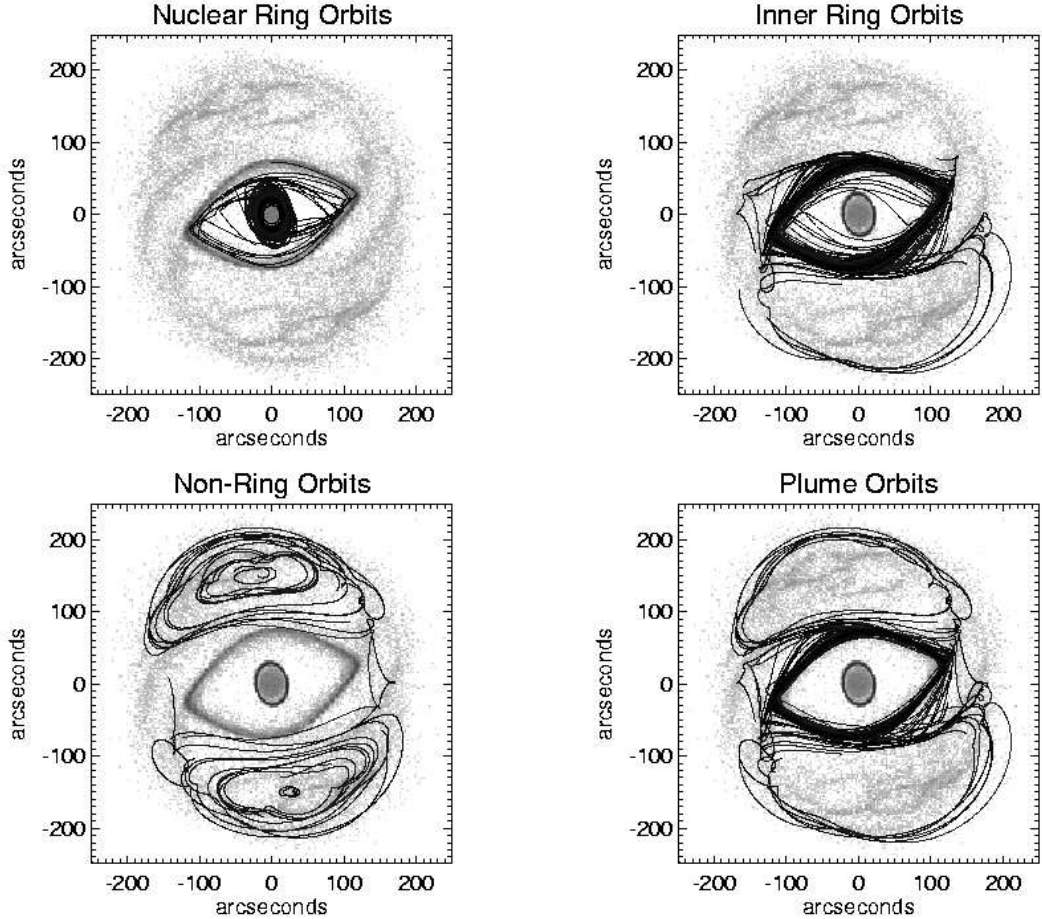}
\label{sample}
\caption{Plots of selected gas particle orbits from 1.0 to 5.0 bar
rotations overlayed on the simulated gas particle distribution at 5.0 bar
rotations with
$\Omega_p$ = 0.89 km s$^{-1}$ arcsec$^{-1}$. The upper left and upper right
plots display the orbits that conclude in the nuclear and inner ring of the
simulation, respectively. The lower left plot displays orbits trapped near the
$L_{4}$ and $L_{5}$ Lagrangian points. The lower right plot displays orbits
that pass through the regions of the plumes. Most of these gas particles were
found to conclude their orbits in the inner ring. All of the plots are in the
disk plane of the simulation and the orbits are shown in the rotating frame.}
\end{figure}

\end{document}

%% file: tab1.tex
\begin{deluxetable}{llcc}
\tabletypesize{\scriptsize}
\tablewidth{0pc}
\tablecaption{Deprojected Feature Shape and Sizes\tablenotemark{a}}
\tablehead{
\colhead{Feature} 
&\colhead{}
&\colhead{$a$}
&\colhead{$b/a$}\\
\colhead{(1)}
&\colhead{(2)}
&\colhead{(3)}
&\colhead{(4)}
}

\startdata
Nuclear ring     & Model       & 28.4 $\pm$ 2.2 & 0.76 $\pm$ 0.07 \\
                 & Observation & 8.9            & 0.9             \cr\\
Inner ring       & Model       & 113 $\pm$ 8    & 0.56 $\pm$ 0.07 \\ 
                 & Observation & 107            & 0.63            \cr\\
Plumes           & Model       & 167 $\pm$ 12   & 0.63 $\pm$ 0.09 \\ 
                 & Observation & 183            & 0.59            \cr\\
Outer pseudoring & Model       & 176 $\pm$ 15   & 0.98 $\pm$ 0.13 \\ 
                 & Observation & 190            & 0.89            \\
\enddata
\tablenotetext{a}{Explanation of columns: (1) ring or feature; (2) data 
extracted from our best fitting model at 5 bar rotation periods with 
$\Omega_p$ = 0.89 km s$^{-1}$ arcsec$^{-1}$ and observations from Buta et al. 
(2001) where both assume a position angle of 21$^{\circ}$ and an inclination of 33$^{\circ}$; (3) semimajor-axis radius in arcseconds; (4) axis ratio.}
\end{deluxetable}

%% file: tab2.tex
\begin{deluxetable}{lcc}
\tabletypesize{\scriptsize}
\tablewidth{0pc}
\tablecaption{Resonance Locations in NGC 1433}
\tablehead{
\colhead{Resonance} 
&\colhead{Radius\tablenotemark{a}}
&\colhead{Radius\tablenotemark{b}}\\
\colhead{}
&\colhead{(arcsec)}
&\colhead{(arcsec)}
}

\startdata
Inner Lindblad resonance (ILR)            & 30.2{\tablenotemark{c}}      & 54.0 \\
Inner 4:1 ultraharmonic resonance (IUHR)  & 88.9                         & 115.3\\ 
Corotation (CR)                           & 120.9                        & 165.1\\ 
Outer 4:1 ultraharmonic resonance (OUHR)  & 151.1                        & 223.0\\ 
Outer Lindblad resonance (OLR)            & 192.0                        & 274.5\\ 
\enddata
\tablenotetext{a}{Resonance radii from Table 5 of Buta et al. 2001 assuming a distance of 11.6 Mpc.}
\tablenotetext{b}{Resonance radii from our best fitting model, as seen in Figure 5.}
\tablenotetext{c}{The outer inner Lindblad resonance radius is given.}
\end{deluxetable}